\documentclass[twocolumn,superscriptaddress,aps,pra,10pt]{revtex4-1}
\usepackage{amssymb,amsmath,amsthm}
\usepackage{graphicx}
\usepackage{color}
\usepackage{bbold}
\usepackage{epsfig}

\newcommand{\rd}{\mathrm{d}}

\begin{document}
	
	\title{Gaussian state entanglement witnessing through lossy compression}
	
\author{Waldemar K{\l}obus}
\affiliation{Institute of Theoretical Physics and Astrophysics, Faculty of Mathematics, Physics and Informatics, University of Gda\'nsk, 80-308 Gda\'nsk, Poland}

\author{Pawe{\l} Cie\'sli\'nski}
\affiliation{Institute of Theoretical Physics and Astrophysics, Faculty of Mathematics, Physics and Informatics, University of Gda\'nsk, 80-308 Gda\'nsk, Poland}

\author{Lukas Knips}
\affiliation{Max-Planck-Institut f\"{u}r Quantenoptik, Hans-Kopfermann-Stra{\ss}e 1, 85748 Garching, Germany}
\affiliation{Department f\"{u}r Physik, Ludwig-Maximilians-Universit\"{a}t, Schellingstra{\ss}e 4, 80799 M\"{u}nchen, Germany}
\affiliation{Munich Center for Quantum Science and Technology (MCQST), Schellingstra{\ss}e 4, 80799 M\"{u}nchen, Germany}

\author{Pawe{\l} Kurzy\'nski }
\affiliation{Faculty of Physics, A. Mickiewicz University, Uniwersytetu Pozna\'nskiego 2, 61-614 Pozna\'n, Poland}

\author{Wies{\l}aw Laskowski}
\affiliation{Institute of Theoretical Physics and Astrophysics, Faculty of Mathematics, Physics and Informatics, University of Gda\'nsk, 80-308 Gda\'nsk, Poland}
\affiliation{International Centre for Theory of Quantum Technologies, University of Gda\'nsk, 80-308 Gda\'nsk, Poland}

\begin{abstract}
We propose a method to witness entanglement between two continuous-variable systems in a Gaussian state. Its key ingredient is a local lossy state transfer from the original spatially separated systems onto two spatially separated qubits. The qubits are initially in a pure product state, therefore by detecting entanglement between the qubits we witness entanglement between the two original systems. This method greatly simplifies entanglement witnessing in complex systems. 
\end{abstract}
	
	\maketitle

\section{Introduction}

Entanglement is considered one of the key resources in quantum information science \cite{Horodecki}. It naturally emerges in the majority of many-body systems \cite{manybody} and can be engineered on various experimental platforms \cite{entgen1,entgen3,entgen4,entgen5,entgen6,multiphoton}. However, despite the fact that entanglement seems to be all around us, its detection is challenging, especially in high-dimensional and continuous-variable systems. Detection of entanglement often requires a partial tomography of the system's state \cite{edet}, whose full description is determined by a number of measurements which grows exponentially with the dimension (and is infinite in the continuous case). 

In this work we focus on the problem how to extract information about entanglement in the state of a complex bipartite system $A$. The main idea is to pair $A$ with a simple system $B$ \cite{Eberly}. $A$ is assumed to be difficult to analyze, whereas $B$ allows full analysis. In other words, we limit the interaction with $A$ to the minimum, whereas we are allowed to perform full tomography on $B$. The goal is to learn whether $A$ is entangled by studying solely $B$. At this point we stress that the two subsystems $A$ and $B$ can be defined as separate particles, or as different degrees of freedom of a single particle (e.g., path/polarization, time/polarization, etc., see \cite{multiphoton}).

In particular, we propose a method to detect entanglement between two continuous-variable systems in a Gaussian state by transferring their state onto a state of two qubits and then by analyzing the resulting two-qubit state. In order to develop some intuition, we first show how to design a protocol to detect entanglement between two qu$d$its and then we generalize it to the continuous-variable case. The qubits are initially prepared in a separable state. Hence, any entanglement arising between them must stem from the initial entanglement between the more complex systems. 

It is clear that such a state transfer cannot be perfect since the dimension of the system onto which the transfer is made is lower than the dimension of the original system. Therefore, the above process can be considered a lossy compression, which aims to preserve only the relevant information. In this case, we want to keep the information about entanglement and discard anything else.

\section{$d$-level systems}

Before we analyze entanglement and continuous-variable systems, let us first discuss a single qu$d$it ($A$) and a single qubit ($B$). We will introduce a unitary coupling operation which allows to transfer some properties of the system $A$ to the system $B$. Later we will generalize the scheme to a pair: two qu$d$its -- two qubits.

\subsection{Single system}

As coupling operator we use a controlled rotation (CROT), i.e., a rotation of the qubit controlled by the state of the qu$d$it. More precisely, CROT is defined for a bipartite system $AB$ composed of a controlling $d$-level qu$d$it state $A$, and a target qubit $B$, the state of which is rotated along the $y$-axis, by
\begin{eqnarray}\label{crot}
&&U_{\rm CROT} =  \sum_{j=0}^{d-1} |j\rangle\langle j| \otimes \exp\left(-i \sigma_y \xi_j\right) \\
 && =\sum_{j=0}^{d-1} |j\rangle\langle j| \otimes \left(\cos\xi_j \openone -  i \sin \xi_j \sigma_y \right), \nonumber
\end{eqnarray}
where the rotation parameter $\xi_j$ depends on the original state of the qu$d$it, $\xi_j = \frac{j \pi}{2(d-1)}$. After the coupling we ignore the subsystem $A$ by tracing it out, and perform analysis on the qubit $B$. 

The above operation resembles the von Neumann measurement apparatus \cite{vonNeumann}, with the exception that the pointer is not a continuous-variable system, but a single qubit. As a result, the measurement of the observable (in this case $A=\sum_j j|j\rangle\langle j|$) cannot be perfect due to the fact that one can encode at most a single bit of classical information on a single qubit. Nevertheless, we are going to show that after the CROT operation some important information about the qu$d$it's state can be decoded from the qubit's state.

As an example let us consider a $d$-level system being in the state
\begin{equation}
|\psi (p) \rangle_A = \sum_{k=0}^{d-1} \sqrt{{d-1 \choose k} p^k (1-p)^{d-1-k}} ~|k\rangle \label{stated}
\end{equation} 
parametrized by a single unknown parameter $p$.	The probability amplitudes are given by the Bernoulli distribution.
Applying the coupling operation \eqref{crot} on the qu$d$it--qubit pair $|\psi(p)\rangle_A \otimes | 0 \rangle_B $, we get 
\begin{eqnarray}
|\Psi \rangle_{AB} &=& U_{\rm CROT} \left(|\psi(p)\rangle_A \otimes | 0 \rangle_B\right) \nonumber \\
&=& \sum_{j=0}^{d-1}	 \sqrt{{d-1 \choose j} p^j (1-p)^{d-1-j}} ~|j\rangle_A \\
&\otimes& \left(\cos\xi_j |0\rangle_B + \sin\xi_j |1\rangle_B \right). \nonumber
\end{eqnarray} 
The reduced density matrix of the system $B$ is given by	
\begin{eqnarray}
& &\rho_B = {\rm Tr}_A |\Psi\rangle_{AB} \langle \Psi|	 = \sum_{j=0}^{d-1}	 {d-1 \choose j} p^j (1-p)^{d-1-j} \times \nonumber \\
& &\left(
\begin{array}{cc}
\cos^2\xi_j & \frac{1}{2}\sin 2\xi_j \\
\frac{1}{2}\sin 2\xi_j & \sin^2 \xi_j
\end{array}  \right), 
\end{eqnarray}
allowing to extract information about the parameter $p$ by, for example, a measurement along $\sigma_z$, ${\rm Tr} (\rho_B \sigma_z)$.
In the limit of infinite dimensions $d$, 
\begin{equation}
\rho_B \stackrel{d \to \infty}{=}  \left(
\begin{array}{cc}
\cos ^2\left(\frac{\pi  p}{2}\right) & \frac{1}{2} \sin (\pi  p)\\
\frac{1}{2} \sin (\pi  p) & 	\sin ^2\left(\frac{\pi  p}{2}\right)
\end{array}\right),
\end{equation}
$\rho_B$ becomes a pure state $|\Psi\rangle_B = \cos \left(\frac{\pi  p}{2}\right) |0\rangle + \sin \left(\frac{\pi  p}{2}\right) |1\rangle$.  Hence, $p = (1/\pi) \arccos {\rm Tr  (\rho_B \sigma_z) }$.

\subsection{Entangled systems}

Let us now suppose the system $A$ is composed of a pair of $d$-dimensional qu$d$its in the state $|\psi\rangle_A= \sum_{j,l=0}^{d-1} a_{jl} |jl\rangle$, which we want to couple with a pair of qubits $B$. In order to do this, we use the coupling operator $U_{\rm CROT}^{\otimes 2}$ for each pair of subsystems, such that the CROT operator couples the first (second) qu$d$it to its respective qubit. 

If both qubits are initially in the state $|0 \rangle$, then application of $U_{\rm CROT}^{\otimes 2}$ to the total system $|\psi \rangle_A |00\rangle_B$ results in
\begin{eqnarray}
| \Psi\rangle_{AB} &=& U_{\rm CROT}^{\otimes 2}(|\psi \rangle_A \otimes |00\rangle_B) \nonumber \\
&=& \sum_{j,l=0}^{d-1} a_{jl} ~  |j  l \rangle_A  \otimes 	\Big[	\cos\xi_j \cos\xi_l |00\rangle_B \nonumber \\
&& +\cos\xi_j \sin\xi_l |01\rangle_B  +\sin\xi_j \cos\xi_l |10\rangle_B \nonumber \\
&& +\sin\xi_j \sin\xi_l |11\rangle_B  \Big].
\end{eqnarray}
In general, the state $| \Psi\rangle_{AB} $ can be highly four-partite entangled, which results in separable subsystems. Therefore, if we want to transfer entanglement from the system $A$ to $B$, we are obligated to do a conditional (projective) measurement on the system $A$. One of the good candidates is the local projection onto the state $|\!+\!\!+\rangle_A=|+\rangle |+\rangle$ with $|+\rangle  =1/\sqrt{d}\sum_{k=0}^{d-1} |k\rangle$. 
After successful projection, the resulting state reads
\begin{eqnarray}
&& \mathcal{N} |++\rangle_A \sum_{j,l=0}^{d-1}  a_{jl} ~ \Big[ \cos\xi_j \cos\xi_l |00\rangle_B \nonumber \\
&& +\cos\xi_j \sin\xi_l |01\rangle_B  +\sin\xi_j \cos\xi_l |10\rangle_B \nonumber \\
&& +\sin\xi_j \sin\xi_l |11\rangle_B  \Big],
\end{eqnarray}
where $(1/\mathcal{N})^2$ is the probability of projecting the system $A$ of two qu$d$its onto $|+\rangle |+\rangle$.

At this point it is worth to consider an example. Let $A$ be in the maximally entangled state corresponding to $a_{jl}= \delta_{jl}/\sqrt{d}$. Then, after the coupling operation, the overlap of the resulting state $|\Psi\rangle_B$ with the maximally entangled state $\frac{1}{\sqrt{2}}(|00\rangle +  |11\rangle)$ decreases with $d$, 
but asymptotically approaches $\pi^2/(\pi^2 + 4) \approx 0.712$ as $d \to \infty$.

Notice that in the special case of $d=2$, the operation swaps the state of the system $A$ to the system $B$,
\begin{eqnarray}
&&|\Psi \rangle_B \stackrel{d=2}{=} \mathcal{N}^{-1} \sum_{j,l=0}^{1}  a_{jl} ~ \Big[ \cos \left(\frac{j \pi}{2 }\right) \cos \left(\frac{l \pi}{2 }\right) |00\rangle \nonumber \\
&&	+\cos \left(\frac{j \pi}{2 }\right) \sin \left(\frac{l \pi}{2 }\right) |01\rangle   +\sin \left(\frac{j \pi}{2 }\right) \cos \left(\frac{l \pi}{2 }\right) |10\rangle \nonumber \\
&&	+\sin \left(\frac{j \pi}{2 }\right) \sin \left(\frac{l \pi}{2 }\right) |11\rangle  \Big] \nonumber \\
	&&=   a_{00} |00\rangle+ a_{01} |01\rangle + a_{10} |10\rangle  + a_{11}|11\rangle.
	\end{eqnarray}

For a general $d$ the resulting state $|\Psi\rangle_B$ is separable if the input state $|\psi\rangle_A$ is separable. This is because a factorization of the amplitudes $a_{jl}=a_j' a_l''$ allows to factorize the resulting state $|\Psi\rangle_B$,
\begin{eqnarray}
&&| \Psi \rangle_B \stackrel{|\psi_{\rm{prod}}\rangle_A}{=} \sum_{j=0}^{d-1} a_j'(\cos \xi_j |0\rangle + \sin\xi_j|1\rangle) \nonumber \\
&& \otimes \sum_{l=0}^{d-1} a_l''(\cos\xi_l|0\rangle + \sin\xi_l|1\rangle). 
\end{eqnarray}

	
Let us now consider a general mixed state of two qu$d$its, i.e., $\rho_A = \sum_{i,j,k,l=0}^{d-1} \rho_{ij,kl} |i\rangle\langle j|\otimes |k \rangle\langle l|$. If we denote $\rho_{AB}$ as the state of the total system after the coupling,
\begin{eqnarray}
\rho_{AB} = U_{\rm CROT}^{\otimes 2} (\rho_A \otimes |00\rangle_B \langle 00| ) (U_{\rm CROT}^{\otimes 2})^{\dagger},
\end{eqnarray}
projecting the subsystem $A$ onto $|+\!+\rangle$ results in subsystem $B$ becoming
\begin{equation}
\rho_B =  \sum_{i,j,k,l=0}^{d-1} \sum_{m,n,p,q=0}^1 \rho_{ij,kl} \; a^m_i a^n_k a^p_j a^q_l \;  |m\rangle\langle p|\otimes |n \rangle\langle q|,
\end{equation}
where
\begin{equation}
a^{\alpha}_{\beta}=
\begin{cases}
\cos \frac{\beta \pi}{2(d-1)},  {\rm for} \;  \alpha=0, \\
\sin \frac{\beta \pi}{2(d-1)},  {\rm for} \;  \alpha=1.
\end{cases}
\end{equation}
Note that for $d=2$, we have $\rho_B=\rho_A$, the same as in the case of pure states.

Additionally, we also show that a separable state of two qu$d$its is mapped onto a separable state of two qubits. Consider the separable state of two qu$d$its $\rho_A^{\rm{sep}} =\sum_{\lambda}p_{\lambda} \rho^{\lambda}_1 \otimes \rho^{\lambda}_2$, where
\begin{eqnarray}
\rho^{\lambda}_1 &=& \sum_{i,j=0}^{d-1} \rho^{\lambda}_{1,ij} |i \rangle \langle j|, \\
\rho^{\lambda}_2 &=& \sum_{k,l=0}^{d-1} \rho^{\lambda}_{2,kl} |k \rangle \langle l|,
\end{eqnarray}
and hence
\begin{equation}
\rho_A^{\rm{sep}} = \sum_{\lambda}p_{\lambda}  \sum_{i,j,k,l=0}^{d-1} \rho^{\lambda}_{1,ij}\rho^{\lambda}_{2,kl} |i\rangle\langle j|\otimes |k \rangle\langle l|.
\end{equation}
Performing analogous calculations as in the general case for mixed states and taking into account the linearity of all operations, we get a separable state:
\begin{eqnarray}
&&\rho_B \stackrel{\rho_A^{\rm{sep}}}{=}  \sum_{\lambda}p_\lambda \nonumber \\
&\times &\sum_{i,j,k,l=0}^{d-1} \sum_{m,n,p,q=0}^1 \rho^{\lambda}_{1,ij}\rho^{\lambda}_{2,kl} \; a^m_i a^n_k a^p_j a^q_l \;  |m\rangle\langle p|\otimes |n \rangle\langle q| \nonumber \\
&& = \sum_{\lambda}p_\lambda 
\left( \sum_{i,j=0}^{d-1}\sum_{m,p=0}^1 \rho^{\lambda}_{1,ij} \; a^m_i a^p_j |m\rangle\langle p| \right) \nonumber \\
&& \otimes \left( \sum_{k,l=0}^{d-1}\sum_{n,q=0}^1 \rho^{\lambda}_{2,kl} \; a^n_k a^q_l |n \rangle\langle q| \right) .
\end{eqnarray}

Please note that the condition for $\rho_A$ to be separable so that the resulting $\rho_B$ is also separable is only sufficient, not necessary. There are instances of entangled states $\rho_A$ which are not mapped into entangled $\rho_B$, hence the scheme effectively works as an entanglement witness.

\section{Continous-variable systems}

We will now generalize our scheme to the case in which the system $A$ is being described by a continuous-variable state. In this regard we limit our considerations to the broad family of Gaussian states.

\subsection{Single system}

If the first subsystem has a continuous spectrum, the coupling operator reads
\begin{equation}\label{crotcv}
U_{\rm CROT} = \int_{-\infty}^{\infty} \rd x |x\rangle \langle x| \otimes (\cos x \openone - i \sin x \sigma_y).
\end{equation}
Next, consider a Gaussian state
\begin{equation}
|\psi(\sigma,m)  \rangle_A = \int \rd x \frac{1}{(2\pi \sigma^2)^{1/4}} e^{-\frac{(x-m)^2}{4\sigma^2}} ~|x\rangle
\end{equation} 
specified by two parameters ($\sigma, m$).  After applying the coupling operation to $ |\psi(\sigma,m)\rangle_A \otimes | 0 \rangle_B$,	we get 
\begin{eqnarray}
&&U_{\rm 	CROT} \left( |\psi(\sigma,m)\rangle_A \otimes | 0 \rangle_B \right)\\
&=& \int \rd x ~\frac{1}{(2\pi \sigma^2)^{1/4}} e^{-\frac{(x-m)^2}{4\sigma^2}}  ~|x\rangle \otimes (\cos x ~|0\rangle + \sin x~ |1\rangle ).\nonumber
\end{eqnarray} 
The reduced state of the qubit $B$ is	
\begin{equation}
\rho_B=  \frac{1}{2} \left(\openone +   e^{-2 \sigma^2} \left(
\begin{array}{cc}
\cos 2m&  \sin 2m  \\
\sin 2m  & - \cos 2m
\end{array} \right)\right)
\end{equation}
and can be visualized by a Bloch vector $\vec b$ lying in the $xz$-plane. The vector $\vec b$ makes an angle $2m$ with the $z$-axis and its norm is $e^{-2 \sigma^2}$. 
The parameters of the original Gaussian state can be recovered from a tomography on the qubit. In particular, $\sigma^2 = -(1/4)\ln ||\vec b||^2$ and $m=\operatorname{acot}(b_z/b_x)/2$.

\subsection{Entangled systems}

Now, we consider the lossy entanglement transfer from the bipartite Gaussian state onto the two-qubit state. In order to do this, we use the coupling operator of \eqref{crotcv} for each respective pair of subsystems, $U_{\rm CROT}^{\otimes 2}$, such that the first (second) complicated subsystem interacts with its respective qubit. 

After the coupling operation, we project the system $A$ of the two particles onto a product of Gaussian states
\begin{equation}
|x_1^+ x_2^+(\Gamma)\rangle =\int \rd x_1 \int \rd x_2 \frac{1}{(2\pi \Gamma^2)^{1/2}} e^{-\frac{(x_1^2+x_2^2)}{4\Gamma^2}}   ~|x_1\rangle|x_2\rangle.
\end{equation}
This is an analogy to the projection onto a uniform superposition that we used in the two-qu$d$it case. This time the projection is parametrized by a single parameter $\Gamma$, which corresponds to the standard deviation. Note that in the limit $\Gamma \rightarrow \infty$ the Gaussian function becomes a uniform superposition over the whole space, akin to what was considered in the qu$d$it case. Such a projection can be interpreted as a projection onto a ground state of a harmonic oscillator, for which the parameter $\Gamma$ can be manipulated by the oscillator's frequency. 

As an example we consider two particles in an entangled Gaussian state
\begin{eqnarray}
&& |\psi (\sigma,\Sigma) \rangle_{A} = \\
&&\int_{-\infty}^{\infty} \rd x_1 \int_{-\infty}^{\infty} \rd x_2\frac{1}{(2 \pi \sigma \Sigma)^{1/2}} e^{-\frac{(x_1+x_2)^2}{8\sigma^2}} e^{-\frac{(x_1-x_2)^2}{8\Sigma^2}} ~|x_1\rangle |x_2\rangle. \nonumber
\end{eqnarray}
This state is entangled whenever $\sigma \neq \Sigma$. Since it is a pure state, its entanglement can be measured by the purity of a subsystem, which in this case is given by \cite{GaussP}
\begin{equation}\label{P}
P=\frac{2\sigma\Sigma}{\sigma^2 + \Sigma^2}.
\end{equation}

\begin{figure}
\includegraphics[width=0.5\textwidth]{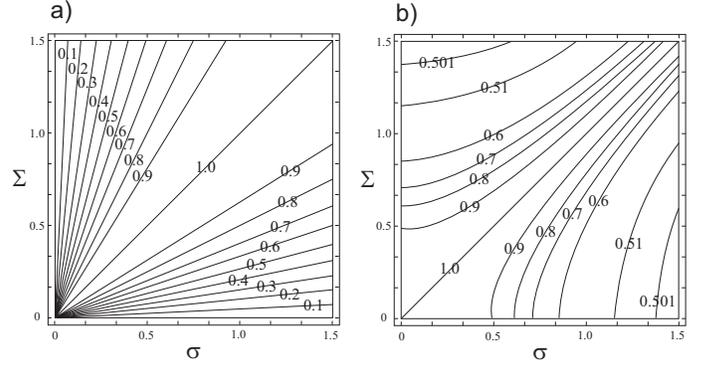}
	\caption{\label{Ss} The purity of a subsystem in a function of $\sigma$ and $\Sigma$ for (a) the original and (b) resulting state.}
\end{figure}

After the coupling operation, the state $|\psi (\sigma,\Sigma) \rangle_{A} \otimes |00\rangle_B$ becomes
\begin{eqnarray}
&& U_{\rm	CROT}^{\otimes 2} (	|\psi (\sigma,\Sigma) \rangle \otimes |00\rangle) \\
&=& \int \rd x_1 \int \rd x_2 \frac{1}{(2 \pi \sigma \Sigma)^{1/2}} e^{-\frac{(x_1+x_2)^2}{8\sigma^2}} e^{-\frac{(x_1-x_2)^2}{8\Sigma^2}} ~|x_1\rangle |x_2\rangle \nonumber\\ &\otimes& 	
\Big[
\cos x_1 \cos x_2 |00\rangle  +\cos x_1 \sin x_2 |01\rangle  \nonumber \\
& +& \sin x_1 \cos x_2|10\rangle  	+\sin x_1 \sin x_2 |11\rangle  \Big]. \nonumber
\end{eqnarray}
After projecting the system $A$ onto $|x_1^+ x_2^+(\Gamma)\rangle$, the state of system $B$ becomes
\begin{equation}
\mathcal{N} (a_+ |00\rangle + a_- |11\rangle),
\end{equation}
where
\begin{equation}
 a_{\pm}=\frac{\pm e^{-\frac{2 \sigma ^2 \Gamma^2}{\sigma ^2+\Gamma^2}}+e^{-\frac{2 \Sigma ^2 \Gamma^2}{\Sigma ^2+\Gamma^2}}}{\sqrt{\frac{\left(\sigma ^2+\Gamma^2\right) \left(\Sigma ^2+\Gamma^2\right)}{\sigma  \Sigma  \Gamma^2}}}.
\end{equation}
The purity of the qubit subsystem $B$ is   
\begin{equation}
P_q = \frac{1}{2} \left(\text{sech}^2\left(\frac{2 \Gamma^4 (\sigma -\Sigma ) (\sigma +\Sigma )}{\left(\sigma ^2+\Gamma^2\right) \left(\Sigma ^2+\Gamma^2\right)}\right)+1\right),
\end{equation}
which becomes in the limit of $\Gamma \to \infty$
\begin{equation}\label{Pql}
\lim_{\Gamma \to \infty} P_q = \frac{1}{2} \left(\text{sech}^2(2 (\sigma -\Sigma ) (\sigma +\Sigma ))+1\right).
\end{equation}
Comparing (\ref{P}) and (\ref{Pql}), we see that in both cases the states are entangled for $\Sigma \neq \sigma$. Also, in both cases the situation $\sigma-\Sigma \to \infty$ corresponds to the maximally entangled states (see Fig. \ref{Ss}).

\begin{figure}
	\includegraphics[width=0.45\textwidth]{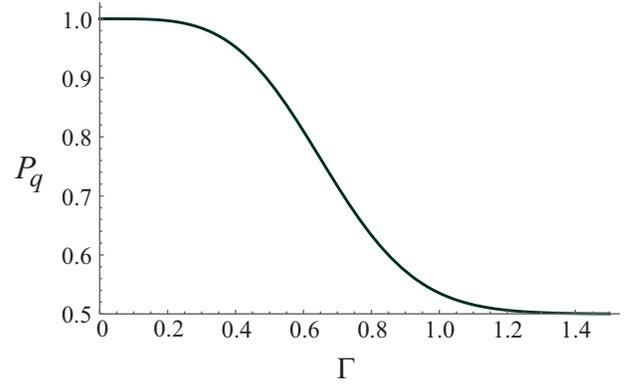}
	\caption{\label{SG} The purity of a subsystem for the resulting state with $\sigma-\Sigma \to \infty$ in a function of the projection parameter $\Gamma$.}
\end{figure}

In Fig. \ref{SG} we present how the purity of a subsystem depends on $\Gamma$. We analyze the extreme case $(\sigma - \Sigma) \to \infty$. In this case the purity is $\frac{1}{2} ({\rm sech}^2 (2 \Gamma^2)+1)$ and decreases with $\Gamma$. Already for $\Gamma$ above $1$, the purity is close to $1/2$.

\subsection{Possible realization}

Here, we give examples for possible implementations of the above scheme. The first concept is in a sense an inverted Stern-Gerlach scenario. We focus on a single system, since the entangled case is a straight-forward generalization.

Consider a spin-$1/2$ particle, say a silver atom, propagating along the z-axis.
The continuous-variable state of interest is encoded in the transversal degree of freedom, say, a spatial state along the $x$-axis $|\psi(x)\rangle$. The spin of the particle is initially pointing up along the z-axis. In addition, consider a region in which there is nonzero magnetic field $\vec B = B(x) \hat y$ pointing along the y-axis with a gradient along the x-axis. We assume that in the region in which $|\psi(x)\rangle$ is supported one can use the approximation $B(x) \approx B_0 x$. This magnetic field region starts at $z=z_0$ and ends at $z=z_1$ $(0<z_0<z_1)$. Outside of this region there is no magnetic field.
The particle starts at $z=0$ and moves towards the magnetic field region with velocity $v$. It spends the time $t=(z_1-z_0)/v$ within the magnetic field region. The magnetic field causes a position dependent rotation of spin about the y-axis 
\begin{equation}
|\uparrow_z\rangle \rightarrow \cos \alpha(x) |\uparrow_z\rangle + \sin \alpha(x) |\downarrow_z\rangle,
\end{equation}
where $\alpha(x) \propto B_0 x (z_1-z_0)/v$. This conditional rotation can be associated with the CROT operation. This way the state $|\psi(x)\rangle$ is lossy transferred onto the spin state.

Another possible implementation is the interaction of different degrees of freedom of photons. 
A natural choice for the degree of freedom of the $d$-level system is using path-encoding as it easily allows to manipulate, say, the polarization state of the photon depending on the path state using waveplates for building up the CROT operation as given in Eq.~(\ref{crot}).
This general concept can be combined with a plethora of different degrees of freedom.
Wavelength division multiplexers allow coupling frequency-bin encoded qu$d$its to qubits using this scheme.
Similarly, when using, e.g., orbital angular momentum (OAM) for the qu$d$it-system, the OAM-encoding can first be translated to path-encoding using a mode sorter~\cite{ModeSorter}.
Recently, a controlled-$\hat X$ gate between the radial degree of freedom of light and its OAM has been shown~\cite{OAMcoupling}, providing another perfect testbed for our coupling.

In a recent work, a high-finesse cavity has been used to couple a ${}^{87}$Rb atom to the coherent state of a light field reflected at the cavity for creating Schr\"odinger cat states~\cite{SchroedingerCat}. 
This technique may also allow to couple the state of the light field to the atom using the CROT operation as given in Eq.~(\ref{crotcv}).
Our proposal could hence facilitate probing for entanglement of two light fields.

Furthermore, our proposal can also be used if both systems $A$ and $B$ are actually qu$d$its.
For example, entanglement of a system of two high-dimensional trapped ions could be probed by properly designing the interaction with two other ions using a suitable modified CROT operation such that the qu$d$its of system $B$ make use of only a two-level submanifold of the ion. This greatly simplifies their read-out as they can now be treated as qubits.
This procedure is applicable also to other high-dimensional systems such as, say, superconducting transmon qu$d$its.

Finally, we would like to mention that our approach also works for multiqubit systems, in which the entanglement between two specific subsets of particles is to be analyzed. The entanglement between a set $A_1$ of qubits and a set $A_2$ of qubits can be studied by first compressing the multiqubit states $\rho_{A_1}$ and $\rho_{A_2}$ into the single qubits $B_1$ and $B_2$, respectively, using a CROT operation. Afterwards, the verification of entanglement of those two single qubit systems implies entanglement between the initial multiqubit systems.

\section{Conclusions}

In this paper we address the problem of detecting entanglement properties of a complex system by analyzing an auxiliary system coupled to the original one. In order to do this we define a coupling operator which transforms the auxiliary system so that after the operation the measured properties of the coupled system provide relevant information about the nature of the original one. Since the auxiliary system is chosen to be of lower dimensionality than the original one, the transfer of information through the coupling operator cannot be exact, hence we can consider the operation a lossy compression. In the process, however, we are being offset by the reduction of the number of measurements required to analyze the entanglement properties of the measured system. Moreover, the scheme works also when we intend to detect entanglement between two continuous-variable systems in a Gaussian state, which in principle can be partially encoded in a simple two-qubit state.  

\section*{Acknowledgments}

W.K., P.K., L.K. and W.L. acknowledge the support by DFG (Germany)
and NCN (Poland) within the joint funding initiative
“Beethoven2” (Grant No. 2016/23/G/ST2/04273).
W.L. acknowledges partial support by the Foundation
for Polish Science (IRAP project, ICTQT, Contract
No. 2018/MAB/5, cofinanced by EU via Smart Growth
Operational Programme).
L.K. is grateful to Marcus Huber and Stephan Welte for helpful comments.


\end{document}